\documentclass[12pt,a4paper]{article}
\usepackage{exscale}
\usepackage{relsize}
\usepackage{amsmath}
\usepackage{amssymb}
\usepackage{amsfonts}
\usepackage{float}
\usepackage{cite}
\usepackage{color}
\usepackage[T1]{fontenc}
\usepackage[utf8]{inputenc}
\usepackage{authblk}
\usepackage{hyperref}
\usepackage{fancyhdr}
\begin{document}
\renewcommand{\thefootnote}{\fnsymbol{footnote}} 
\newcommand{\emp}[1]{{\texttt{#1}}}
\newcommand{\red}[1]{{\color{red} #1}}
\newcommand{\blue}[1]{{\color{blue} #1}}
\newcommand{\yel}[1]{{\color{yellow} #1}}
\newcommand{\ross}[1]{{\color{rossoCP3} #1}}

\begin{center}
\LARGE \rm Entropy of Vaidya Black Hole on Apparent Horizon with Minimal length Revisited \footnote{Supported by National Natural Science Foundation of China under Grant Nos. 11675139, 11605137, 11435006, 11405130 and the Double First-Class University Construction Project of Northwest University. Bin Wu is also supported by the China Postdoctoral Science Foundation (No.2017M623219).}  
\end{center}

\begin{center}
Hao Tang,$^{1,3}$\ \
Bin Wu,$^{1,2,3}$ \footnote{Corresponding author. Email: binwu@nwu.edu.cn.}  \
Cheng-yi Sun,$^{1,3}$  \  \ \\
Yu Song,$^{1,3}$\  \
Rui-hong Yue$^{1,4}$\ \
\end{center}

\begin{center}
\begin{footnotesize}
$^{1}$ Institute of Modern Physics, Northwest University, Xi'an 710069, China\\   
$^{2}$ Shaanxi Key Laboratory for Theoretical Physics Frontiers, Northwest University, Xi'an 710069, China\\  
$^{3}$ School of Physics, Northwest University, Xi'an 710069, China\\  
$^{4}$ College of Physical and Technology, Yangzhou University , Yangzhou 225009, China\\   
\end{footnotesize}
\end{center}

\date{}

\begin{abstract}
By considering the generalized uncertainty principle, the degrees of freedom near the apparent horizon of Vaidya black hole are calculated with thin film model. The result shows that a cut-off can be introduced naturally rather than taking by hand. Furthermore, if the minimal length is chosen to be a specific value, the statistical entropy will satisfy the conventional
area law at the horizon, which might reveal some deep things of the minimal length.
\end{abstract}

\textbf{Keywords:} entropy; black hole; generalized uncertainty principle.
\\
\\
\textbf{PACS numbers:} 04.70.Dy, 97.60.Lf
\\
\\

Black hole is one of the most intriguing objects which possibly provide a bridge between an underlying quantum gravity theory and the classical gravity.
Pioneering by the Bekenstein's \cite{Bekenstein} and Hawking's \cite{Hawking}
works, the black holes could have the temperature and entropy as an ordinary thermodynamical macroscopic system. Moreover,
the entropy of a black hole is argued to be proportional to the area $A$ of the horizon. Since the origin of the entropy of the black hole at the microscopic level becomes an active and fascinating field of researches,
many efforts have been made and several different methods have been developed to calculate the entropy \cite{Many-Methods-01,Many-Methods-02,Many-Methods-03} of black holes at statistical level. In Ref.\cite{brick-wall-thooft}, 't Hooft proposed a widely-used method called ``brick-wall'' by supposing that the entropy of the black hole comes from the contributions of the quantum gases in a ``wall" outside the horizon.
However, there exists divergence of the entropy if no further constraint has been introduced on the brick-wall. So that a cut-off of the brick-wall must be introduced by hand, which is unnatural in some sense\cite{brick-wall-thooft,brick-wall-01,brick-wall-02,brick-wall-03,brick-wall-04,brick-wall-05,brick-wall-06,brick-wall-07,HUP-result-01,HUP-result-02}.

Recently, the generalized uncertainty principle (GUP), which is developed from the Heisenberg's uncertainty principle (HUP),  is used in the brick-wall model to calculate the free energy and the entropy. It shows that the cut-off is naturally given by a minimal length indicated by the GUP \cite{gup,Chang:2001bm,gup-x-01,gup-x-02,gup-x-03,gup-x-04,gup-x-05,gup-x-06,gup-discussions-01,gup-discussions-02,gup-discussions-03,gup-discussions-04}. The bound of the entropy has also been obtained in the framework of the GUP. However, all those works are concentrated on the static spherically symmetric black holes. In this paper,
we are interested in the dynamical spacetime. We will follow the approach of brick-wall method with the GUP to calculate the statistical entropy of Vaidya black hole. Planck unit $c=G=\hbar=k_B=1$ is used throughout this paper for convenience.

Vaidya spacetime is a solution of the Einstein's equation, which describes the formation of a Schwarzschild-like black hole out of a collapsing shell of null dust. The metric of the Vaidya spacetime in the advanced Eddington coordinates reads \cite{Vaidya-Bonner-GUP,Vaidya-AH-EH}
\begin{align}\label{eq:metricAH}
ds^2=-[1-\frac{2m(v)}{r}] {\rm d} v^2 + 2 {\rm d} v {\rm d} r + r^2 {\rm d} \theta^2 + r^2 \sin^2 \theta {\rm d} \phi^2,
\end{align}
where $m(v)$ is the dynamical mass of the black hole, $v$ is the advanced Eddington time coordinate and
\begin{align}\label{eq:fAH}
f(r)=1-\frac{2m(v)}{r}.
\end{align}
In order to characterize the formation of a dynamical black hole,
there is a generalized notion called apparent horizon
which is distinct from the event horizon.
To demonstrate, we just consider the apparent horizon which is a
closed space-like hypersurface of co-dimension two and is located at\cite{Vaidya-AH-EH}
\begin{align}\label{eq:rAH}
r_{AH}=2m(v).
\end{align}
The surface gravity of the apparent horizon is
\begin{align}\label{eq:kappaAH}
\kappa_{AH}=\frac{1}{2 r_{AH}}=\frac{1}{4m(v)}.
\end{align}
The corresponding Hawking temperature is
\begin{align}\label{eq:tempertureAH}
T_{AH}=\frac{\kappa_{AH}}{2 \pi}=\frac{1}{8 \pi m(v)}=\frac{1}{\beta_{AH}},
\end{align}
thus we have
\begin{align}\label{eq:betaAH}
\beta_{AH}=\frac{2 \pi}{\kappa_{AH}}=4 \pi r_{AH}.
\end{align}
The metric function $f(r)$ can be expanded near the horizon $r_{AH}$ as
\begin{align}\label{eq:frAH}
f(r)\approx f'(r_{AH})(r-r_{AH})=2 \kappa_{AH} (r-r_{AH}).
\end{align}

There are several ways to generalize the Heisenberg's uncertainty principle, and the simplest one among of them is \cite{gup,Chang:2001bm,gup-x-01,gup-x-02,gup-x-03,gup-x-06}
\begin{equation}\label{eq:GUP-formulation}
\Delta x \Delta p \geq \hbar + \frac{\lambda}{\hbar}(\Delta p)^2,
\end{equation}
where $\hbar$ is the Planck constant, $\Delta x$ and $\Delta p$ are the spatial coordinates and momentum respectively, and $\lambda$ is a constant at order of the Planck length\cite{gup}. The Eq.\eqref{eq:GUP-formulation} indicate that there exists a minimal length \cite{gup-x-04,gup-x-05}
\begin{equation}\label{eq:deltax-lambda}
\Delta x \geq 2\sqrt{\lambda}.
\end{equation}

Since we are considering a system of quantum boson gas
in a ``wall'' outside the
horizon, the Klein-Gordon equation for a scalar field is introduced
\begin{equation}\label{equ:KG-function}
(\Box -\mu^2)\Phi=0.
\end{equation}
Decompose the wave function into the form of
\begin{equation}
\Phi=e^{-i\omega v}\psi(r,\theta,\phi),
\end{equation}
and substitute it into Eq.\eqref{equ:KG-function}, yields
\begin{equation}
\begin{split}
\frac{\partial^2\psi}{\partial r^2}+(\frac{2}{r}+\frac{d f(r) }{f(r) dr})\frac{\partial\psi}{\partial r} + \frac{1}{f(r)}[\frac{\omega^2}{f(r)}-\mu^2+ \\
\frac{1}{r^2}(\frac{\partial^2}{\partial \theta^2} + \frac{\partial}{\tan \theta \partial \theta} + \frac{\partial}{\cos^2 \theta \partial \phi^2})]\psi=0.
\end{split}
\end{equation}
By using the Wenzel-Kramers-Brillouin (WKB) approximation, i.e., take $\psi \sim e^{i\mathbb{S}(r,\theta,\phi)}$, one could get \cite{gup-method}
\begin{equation}
p_r^2=\frac{1}{f(r)}[\frac{\omega^2}{f(r)} - \mu^2 - \frac{p_\theta^2}{r^2} - \frac{p_\phi^2}{r^2 \cos^2 \theta}],
\end{equation}
where
\begin{equation}
p_r=\frac{\partial \mathbb{S}}{\partial r}, \ \ p_\theta=\frac{\partial \mathbb{S}}{\partial \theta}, \ \ p_\phi=\frac{\partial \mathbb{S}}{\partial \phi}.
\end{equation}
And the square module momentum is
\begin{equation}\label{eq:pp}
p^2=p_i p^i = g^{rr}p_r^2 + g^{\theta \theta}p_\theta ^2 + g^{\phi \phi}p_\phi^2 = \frac{\omega^2}{f(r)}-\mu^2.
\end{equation}
Eq.\eqref{eq:pp} implies that $\omega \geqslant \mu \sqrt{f(r)}$.

In the framework of the GUP, the number density now becomes \cite{Chang:2001bm}
\begin{equation}
\frac{d^3x d^3p}{(2 \pi )^3(1+\lambda p^2)^3}.
\end{equation}
Consequently, we obtain the number of quantum states as
\begin{align}\label{equ:nw-density}
n(\omega)&=\frac{1}{(2\pi)^3}\int dr d\theta d\phi dp_r dp_\theta dp_\phi \frac{1}{(1+\lambda p^2)^3} \nonumber \\
&=\frac{2}{3\pi} \int dr \frac{r^2(\omega^2/f(r)-\mu^2)^{3/2}}{\sqrt{f(r)}[1+\lambda(\omega^2/f(r)-\mu^2)]^3}.
\end{align}
One could see that it becomes finite even at the horizons. So that a cut-off by hand is needless with the help of GUP.

The quantum boson gases in a “wall” outside the horizon are regarded as the eigenmodes of the scalar field in the vicinity of the black hole. Every wave solution may be occupied by any integer number of quanta, that the total number of the quantum boson particles may be not conserved. This is somewhat analogous to the arguments on the statistics of the photon gas which leads to the blackbody spectrum. In this case, the parameter $\alpha=-\mu/kT$ in the normal Bose-Einstein distribution disappears, thus the chemical potential vanishes.
Now the free energy of the scalar field reads
\begin{equation}
F=-\int_{\mu \sqrt{f(r)}}^\infty d\omega \frac{n(\omega)}{e^{\beta \omega}-1}.
\end{equation}
Therefore, the entropy can be calculated as
\begin{align}\label{equ:entropy}
S&=\beta^2 \frac{\partial F}{\partial \beta} \nonumber\\
&=\frac{2 \beta^2}{3 \pi} \int dr \frac{r^2}{\sqrt{f(r)}} \int_{\mu \sqrt{f(r)}}^\infty d\omega \frac{\omega (\omega^2/f(r) - \mu^2)^{3/2}}{4 \sinh^2(\beta \omega/ 2)\,[1+\lambda(\omega^2/f(r) - \mu^2)]^3} \nonumber\\
&=\frac{\beta^3}{12 \pi \lambda^3} \int r^2 f(r) dr \int_{x_0}^\infty dx \frac{x (x^2 - x_0^2)^{3/2}}{\sinh^2 x \,[x^2-x_0^2+\beta^2 f(r)/(4\lambda)]^3},
\end{align}
where $x \equiv \beta \omega/ 2$ and $x_0=\beta \mu \sqrt{f(r)}/2$.

It is supposed that the main contribution of entropy comes from a small region just near the horizon \cite{LZ-MEM}, whose thickness corresponds to a proper distance at the order of the minimal length indicated by the GUP. Now we consider a thin layer located between $r_{AH}$ and $r_{AH}+\epsilon_{AH}$. Then the range of integration in Eq.\eqref{equ:nw-density} becomes $r_{AH}$ to $r_{AH}+\epsilon_{AH}$ \cite{YXJ-ZZ-01}. This method is usually called as the thin film model, which is also developed from the brick-wall model with GUP. Notice that $x_0$ in the above equation goes to zero at the horizon, the entropy becomes
\begin{align}\label{eq:SH}
S=\frac{\beta^3}{12 \pi \lambda^3}\int_0^\infty dx\frac{x^4}{\sinh^2x}I(x)_{AH},
\end{align}
where
\begin{align}\label{eq:Ix}
I(x)_{AH}=\int_{r_{AH}}^{r_{AH}+\epsilon_{AH}}dr\frac{r^2 f(r)}{[x^2+\beta^2f(r)/(4\lambda)]^3}.
\end{align}

On the other hand, the minimal length can be extracted from the relation
as below
\begin{align}
2\sqrt{\lambda}=\int_{r_{AH}}^{r_{AH}+\epsilon_{AH}}\frac{dr}{\sqrt{f}}\approx\int_{r_{AH}}^{r_{AH}+\epsilon_{AH}}\frac{dr}{\sqrt{2 \kappa_{AH} (r-r_{AH})}}=\sqrt{\frac{2 \epsilon_{AH}}{\kappa_{AH}}},
\end{align}
which gives
\begin{align}\label{eq:epsilonAH}
\epsilon_{AH}=2 \lambda \kappa_{AH}.
\end{align}
Eq.\eqref{eq:Ix} on apparent horizon of Vaidya black hole now becomes
\begin{align}\label{eq:IxAH}
I(x)_{AH}&=\int_{r_{AH}}^{r_{AH}+\epsilon_{AH}} dr \frac{f(r) r}{[x^2+ \frac{f(r) \beta_{AH}}{4 \lambda}]^3} \nonumber \\
&\approx\frac{f'(r_{AH}) \epsilon_{AH}^2 r_{AH}^2}{2 x^2[x^2+\frac{\beta_{AH}^2 f'(r_{AH}) \epsilon_{AH}}{4 \lambda}]^2} \nonumber \\
&=\frac{4 \kappa_{AH}^3 r_{AH}^2 \lambda^2}{(4 \pi^2 x + x^3)^2}.
\end{align}
where $x = \frac{\beta_{AH}\omega}{2}$. Then the entropy is
\begin{align}\label{eq:S-AH}
S_{AH}&=\frac{\beta_{AH}^3}{12 \pi \lambda^3}\int_0^\infty dx \frac{x^4}{\sinh ^2 x} I(x)_{AH} \nonumber \\
&=\frac{A_{AH}}{4} \frac{8\pi}{3\lambda}\int_0^\infty dx \frac{x^4}{\sinh ^2 x (4 \pi^2 x+x^3)^2},
\end{align}
where $A_{AH}=4\pi r_{AH}^2$ is the surface area of the apparent horizon. The integrand in Eq.\eqref{eq:S-AH} could be regarded as a complex function $h(z)=\frac{z^4}{\sinh^2 z (4 \pi^2 z + z^3)^2}$. With the help of residue theorem, Eq.\eqref{eq:S-AH} turns out to be
\begin{align}\label{eq:S-AH-final}
S_{AH}&=\frac{A_{AH}}{4} \frac{8}{3\lambda} (-\frac{1}{24}+\frac{1}{\pi^2} \sum_{n=1, n\neq2, n \in \mathbb{N_+}}^{\infty}\frac{n(n^2+4)}{(n^2-4)^3}) \nonumber \\
&=\frac{A_{AH}}{4} \frac{8}{3\lambda}(-\frac{1}{24}-\frac{25}{32\pi}+\frac{\zeta(3)}{\pi}) \nonumber \\
&=\frac{A_{AH}}{4} \frac{\sigma}{\lambda},
\end{align}
where $\sigma= \frac{8}{3}(-\frac{1}{24}-\frac{25}{32\pi}+\frac{\zeta(3)}{\pi})$. We could see that the entropy of the Vaidya black hole on apparent horizon is proportional to the surface area at the horizon.

If we further require the entropy satisfy the area law of black hole, one need to make $\sigma = \lambda$. Then the minimal length could be obtained as $2\sqrt{\lambda}=2\sqrt{\sigma}l_p\approx 0.254969 l_p$, where $l_p$ is the Planck length.
This requirement might provide some deep insight into the minimal length
indicated by GUP. Since the entropy is consider as the
semi-classical property of Hawking radiation from the horizon within a thin layer of Planck order, so that the range of integration in our method is more reasonable than brick-wall model with a man-made cut-off.


In summary, we have calculated the entropy of apparent horizon with thin film model based on generalized uncertainty principle. We have found that the entropy is proportional to the surface area at the horizon, which is consistent with the results of Bekenstein and Hawking. Our result is more reasonable and reliable because there is no unnatural cut-off in this method. Furthermore, the area law of the black hole
requires the minimal length $\lambda$ in GUP to be a specific value, which deserves future studies to provide some deep insight into microscopic structure of the black hole.

\end{document}